\documentclass[pre,aps,amssymb,amsmath,superscriptaddress,twocolumn, nofootinbib]{revtex4}
\usepackage{amsmath}
\usepackage{amssymb}
\usepackage{epsfig}
\usepackage{amscd}
\usepackage{graphicx,psfrag,xspace}
\usepackage{float}
\usepackage[normalem]{ulem}

\begin{document}

\title{A CTRW approach to normal and anomalous reaction--diffusion processes}
\author{A. Zoia}
\email{andrea.zoia@polimi.it}
\affiliation{E. Fermi Center for Nuclear Studies, Energy Department, Polytechnic of Milan, Milan 20133, Italy}


\begin{abstract}
We study the dynamics of a radioactive species flowing through a porous material, within the Continuous-Time Random Walk (CTRW) approach to the modelling of stochastic transport processes. Emphasis is given to the case where radioactive decay is coupled to anomalous diffusion in locally heterogeneous media, such as porous sediments or fractured rocks. In this framework, we derive the distribution of the number of jumps each particle can perform before a decay event. On the basis of the obtained results, we compute the moments of the cumulative particle distribution, which can be then used to quantify the overall displacement and spread of the contaminant species.
\end{abstract}
\maketitle

\section{Introduction}

The investigation of transport processes in inhomogeneous geological formations has attracted intense research efforts, because of its relevance in the context of subsurface waste management and environmental remediation \cite{dagan, demarsily, rev_geo, sahimi, gelhar}. In such complex physical systems, the spread of the transported quantity is often experimentally found to exhibit a non-linear growth with respect to time, of the kind $\left\langle x^2(t)\right\rangle \sim t^\gamma$, $\gamma \ne 1$. This scaling is actually the hallmark of the so-called anomalous diffusion, as opposed to Fickian (normal) diffusion, where $\gamma=1$ \cite{klafter_sokolov, rev_geo}.

The migration of contaminant particles through both homogeneous and heterogeneous materials has been successfully described within the Continuous-Time Random Walk (CTRW) scheme. In this stochastic model, the trajectory of each particle is represented as a series of random jumps separated by random waiting times, during which the walker stays at rest in the previously reached position \cite{weiss, berkowitz_scher, klafter1, klafter2, rev_geo}. For sake of simplicity, we adopt the common assumption that jumps and waiting times are independent of each other \cite{rev_geo, cortis_berkowitz}. The jump lengths are usually drawn from a Gaussian distribution with (finite) variance $\sigma^2$, where $\sigma$ is a typical spatial scale depending on the traversed material, and mean $\mu$ \cite{cortis_berkowitz, rev_geo}. A forward bias $\mu>0$ is often used to model the contribution of an external advection field  \cite{cortis_berkowitz}. In the context of underground particle flow through porous sediments or bedrock, the migrating plume is most frequently characterized by an anomalous spread ($\gamma \ne 1$), induced by the presence of, for instance, dead ends, stagnation and obstacles, which affect the particle dynamics at the microscopic scale \cite{berkowitz_klafter, berkowitz_scher2, cortis_gallo, marseguerra_zoia, zoia}. These processes are mirrored in extremely long trapping times, which, within the CTRW formulation, are modelled by assuming that the waiting time distribution has a power-law decay $w(t) \sim t^{-1-\alpha}$ with $0<\alpha<2$ \cite{rev_geo, berkowitz_scher, klafter1}. The broad distribution of spatial length scales which characterizes heterogeneous materials can result in a broad (power-law) distribution of characteristic time scales, so that extreme events, i.e. anomalously long resting times, have a non negligible probability of being sampled. This phenomenological picture is at the basis of the CTRW formulation \cite{rev_geo, cortis_berkowitz, berkowitz_scher3, berkowitz_scher}.

In the case of independent jumps and waiting times, the general form of the CTRW transport equation for the normalized particle concentration $P(x,t)$ can be expressed as follows
\begin{equation}
\frac{\partial}{\partial t}P(x,t) = {\cal M} \left( \frac{\sigma^2}{2}\frac{\partial^2}{\partial x^2} -\mu \frac{\partial}{\partial x} \right) P(x,t),
\label{fde}
\end{equation}
where the time convolution operator ${\cal M}$, with a kernel $M(t)$, takes into account possible non-Markovian (memory) effects due to power-law waiting times (see Appendix \ref{appendix} for details) \cite{rev_geo, cortis_berkowitz}. In particular, one-dimensional transport with a constant bias is subdiffusive ($\gamma<1$) when $0<\alpha<1/2$ and superdiffusive ($\gamma>1$) when $1/2<\alpha<2$, as shown, e.g., in \cite{shlesinger, margolin, dentz_berkowitz}.

On the other hand, transport in a locally homogeneous material can be described by assuming that the asymptotic decay of $w(t)$ is sufficiently fast (as it is the case of an exponential distribution), so that the particles wait on average the same characteristic time between any successive jumps \cite{klafter1, weiss, berkowitz_scher}. In this case, Eq. (\ref{fde}) reduces to the well-known normal advection--diffusion equation \cite{klafter1, rev_geo, cortis_berkowitz}. Note that the general formalism of CTRW can account also for a transition from anomalous to normal diffusion, by adopting for instance a truncated power-law distribution with an exponential cut-off for the waiting times: this behavior is often observed in contaminant migration (see, e.g., \cite{rev_geo, cortis_berkowitz, dentz, blunt}).

The theoretical framework of CTRW is well established and has been corroborated by a huge amount of experimental evidences \cite{rev_geo, cortis_berkowitz, berkowitz_kosakowski, berkowitz_scher3, levy_berkowitz, kimmich, note1}. However, due to the subtleties involved in the non-Markovian nature of the memory kernel \cite{shushin, fedotov, rev_geo}, much ingenuity has been necessary to couple reaction phenomena with anomalous diffusion \cite{sokolov_decay1, sokolov_decay2, deem1, deem2}. A comprehensive theoretical treatment, though, is still lacking: see, e.g., \cite{sokolov_decay1} and references therein.

\begin{figure}[t]
 \centerline{ \epsfclipon \epsfxsize=8.0cm
\epsfbox{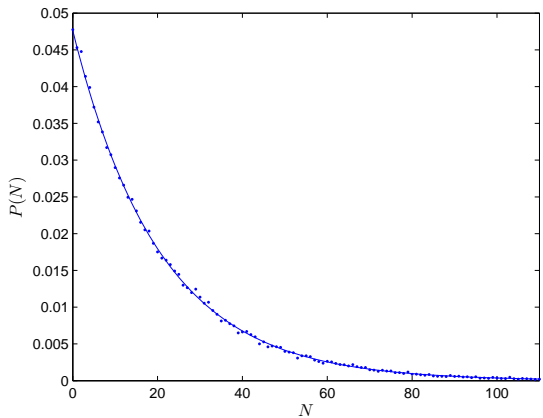} }
   \caption{The distribution ${\cal P}(N)$ (Eq. (\ref{PN_exp}), solid line) is compared with Monte Carlo simulation (dots) for the following parameters: $10^5$ simulated particles, $\tau_0=2$ and $\tau=0.1$.}
   \label{fig1}
\end{figure}

In this paper, we consider the simple but significant case of a system composed of two diffusing species, say $m$ and $n$, where $m$ is unstable and decays through a nuclear reaction to $n$, which is stable. The decay is governed by a Poisson process with parameter $\lambda$. This system can characterize, e.g., the transport of a radioactive contaminant species leaking from an underground repository and migrating through the surrounding geological formations. In analogy with the well known normal reaction--advection--diffusion equations, it would be tempting to postulate a generalization of (\ref{fde}) with a decoupled structure of the kind
\begin{equation}
\frac{\partial}{\partial t}P_j(x,t) = {\cal M}_j{\cal K}_j P_j(x,t) \mp \lambda P_m(x,t),
\label{P_decoupled}
\end{equation}
where ${\cal K}_j = \sigma^2_j \partial_x^2/2 -\mu_j \partial_x$ is the transport operator and  $j=m,n$ \cite{henry}. However, suitably extending the derivation of the CTRW scheme presented in \cite{sokolov_decay1} it is possible to show that the concentrations of $m$ and $n$ obey to
\begin{equation}
\frac{\partial}{\partial t}P_m(x,t) = {\cal M}_m^\ast {\cal K}_m P_m(x,t) - \lambda P_m(x,t)
\label{P_m}
\end{equation}
and
\begin{equation}
\frac{\partial}{\partial t}P_n(x,t) ={\cal M}_{n} {\cal K}_n P_n(x,t) + \lambda P_m(x,t),
\end{equation}
respectively, where the operator
\begin{equation}
{\cal M}_m^\ast = e^{-\lambda t} {\cal M}_{m} e^{\lambda t}
\end{equation}
involves also reaction ($\lambda$) terms \cite{sokolov_decay1, sokolov_decay2}. Thus, only the equation for the species $n$ has a decoupled structure, where transport and reaction act independently. It can be shown that when $w_j(t)$ is an exponential distribution the standard reaction--advection--diffusion equations are recovered, namely
\begin{equation}
\frac{\partial}{\partial t}P_j(x,t) ={\cal T}_j P_j(x,t) \mp \lambda P_m(x,t),
\label{P_diff}
\end{equation}
where ${\cal T}_j = D_j\partial_x^2-v_j\partial_x$ is the transport operator and $D_j$, $v_j$ are the diffusion coefficient and the velocity of each species $j=m,n$, respectively \cite{sokolov_decay1}.

We have implicitly assumed that particles $m$ can still undergo a nuclear reaction when trapped in a stagnant region, and further that particles $n$ once created have different physical-chemical properties from $m$: these represent reasonable hypotheses in the context of radionuclides migration. The concentration profiles corresponding to equations (\ref{P_decoupled}) and (\ref{P_m}), respectively, have been contrasted in \cite{sokolov_decay1}: discrepancies are clearly visible, so that in principle it should be possible to select the proper model on the basis of available experimental data. Other possible implementations of reaction--diffusion phenomena within the CTRW formulation exist (see, e.g., \cite{hornung, sokolov_decay1}), relying on different physical assumptions and thus leading to different transport equations.

Having this framework in mind, in the following we address the issue of computing the number of jumps a diffusing particle $m$ can perform before decaying to $n$, and the corresponding overall displacement and spread of the radioactive species. In Section \ref{jump_decay} we outline the mathematical formalism. Then, in Sections \ref{normal} and \ref{subdiffusion} we discuss the case of normal and anomalous diffusion, respectively. Conclusions are finally drawn in Section \ref{conclusions}.

\section{Number of jumps before decay}
\label{jump_decay}

\begin{figure}[t]
 \centerline{ \epsfclipon \epsfxsize=8.0cm
\epsfbox{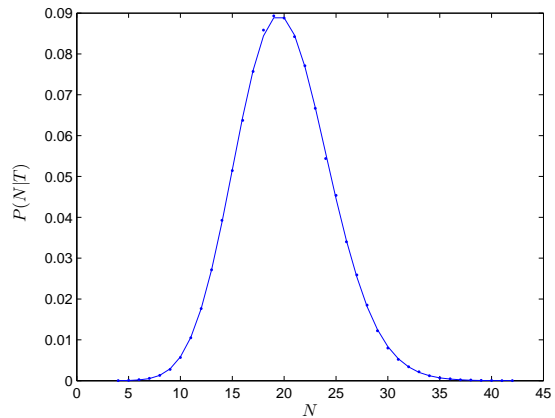} }
   \caption{The distribution ${\cal P}(N|T)$ (Eq. (\ref{PNT_exp}), solid line) is compared with Monte Carlo simulation (dots) for the following parameters: $10^5$ simulated particles, $T=2$ and $\tau=0.1$.}
   \label{fig2}
\end{figure}

Assume that the waiting times between consecutive jumps are sampled from independent and identically distributed probability density functions (pdf's) $w(t)$. Let $\bar{w}(u)={\cal L}\left\lbrace w(t) \right\rbrace$ denote the Laplace transform of $w(t)$. Then, the distribution $w_N(t)$ of $t$ after $N$ jumps will be given by the $N$-fold convolution of $w(t)$ with itself: in the transformed space, we simply have $\bar{w}_N(u)=\bar{w}(u)^N$. Define ${\cal P}(N|T)$ as the probability that a particle whose waiting times are distributed according to $w(t)$ performs $N$ jumps before a final time $T$. The basic relation between the counting process ${\cal P}(N|T)$ and the pdf $w(t)$ of the waiting times between consecutive events is
\begin{equation}
{\cal P}(N|T)=W_N(T)-W_{N+1}(T),
\end{equation}
where $W_N(T)$ is the cumulative distribution associated to $w_N(t)$, evaluated at $T$ \cite{feller}. In Laplace space, ${\cal P}(N|u)=u^{-1}(\bar{w}_N(u)- \bar{w}_{N+1}(u))=u^{-1}(\bar{w}(u)^N-\bar{w}(u)^{N+1})$. Therefore we have
\begin{equation}
{\cal P}(N|T)={\cal L}^{-1}\left\lbrace \frac{1}{u}\left(\bar{w}(u)^N-\bar{w}(u)^{N+1}\right) \right\rbrace(T).
\label{PNT}
\end{equation}

\begin{figure}[t]
 \centerline{ \epsfclipon \epsfxsize=8.0cm
\epsfbox{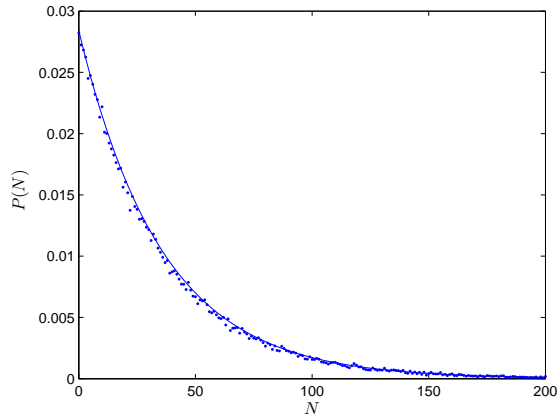} }
   \caption{The distribution ${\cal P}(N)$ (Eq. (\ref{PN_pow}), solid line) is compared with Monte Carlo simulation (dots) for the following parameters: $10^5$ simulated particles, $\alpha=0.5$, $\tau_0=4$ and $\tau=10^{-3}$.}
   \label{fig3}
\end{figure}

Let now $f(T)=\lambda e^{-\lambda T}$ be the pdf of the radioactive decay events. Then, the probability that particles $m$ perform $N$ jumps before decaying to $n$ is
\begin{equation}
{\cal P}(N)= \int_0^\infty {\cal P}(N|T) f(T) dT.
\end{equation}
Integrating once by parts we get
\begin{eqnarray}
{\cal P}(N)= \int_0^\infty e^{-\lambda T} {\cal L}^{-1}\left\lbrace \bar{w}(u)^N \right\rbrace(T) dT + \nonumber \\
- \int_0^\infty e^{-\lambda T} {\cal L}^{-1}\left\lbrace \bar{w}(u)^{N+1} \right\rbrace(T) dT.
\end{eqnarray}
Thus, interpreting each integral as a Laplace transform evaluated at $u=\lambda$ with respect to the internal argument ${\cal L}^{-1}\left\lbrace \bar{w}(u)^N \right\rbrace(T)$, we finally have
\begin{equation}
{\cal P}(N)= \bar{w}(\lambda)^N -\bar{w}(\lambda)^{N+1}.
\end{equation}
Now, in order to characterize the displacement and the spread of the radioactive species $m$ before decay, we are interested in computing the moments of the cumulative particle distribution $P^c_m(x)$, namely
\begin{equation}
E[x^r]= \lambda \int x^r P^c_m(x) dx,
\end{equation}
where
\begin{equation}
P^c_m(x)=\int_0^{+\infty} P_m(x,t) dt
\end{equation}
and the factor $\lambda$ is used to normalize the moments to the average radionuclide decay time. These quantities can be intuitively expressed in terms of the moments of the particles locations pdf after $N$ jumps, $p_N(x)$, averaged on the distribution ${\cal P}(N)$:
\begin{equation}
E[x^r] = \sum_{N =0}^\infty {\cal P}(N) \int x^r p_N(x) dx,
\label{eq_moments}
\end{equation}

\begin{figure}[t]
 \centerline{ \epsfclipon \epsfxsize=8.0cm
\epsfbox{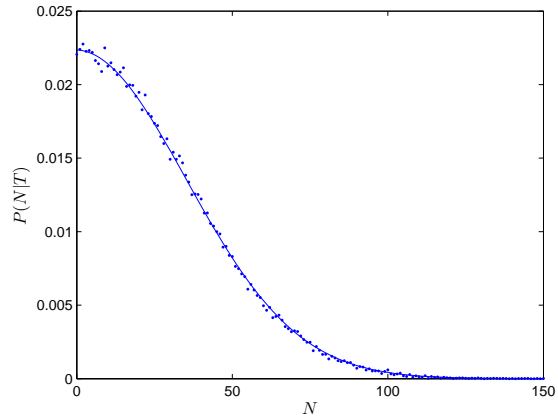} }
   \caption{The distribution ${\cal P}(N|T)$ (Eq. (\ref{PNT_pow}), solid line) is compared with Monte Carlo simulation (dots) for the following parameters: $10^5$ simulated particles, $\alpha=0.5$, $T=2$ and $\tau=10^{-3}$.}
   \label{fig4}
\end{figure}

This can be understood as follows. First, note that, if $Q_m(x,t)$ satisfies Eq. (\ref{fde}) (without radioactive decay), then $P_m(x,t)= Q_m(x,t)e^{-\lambda t}$ satisfies Eq. (\ref{P_m}) for the reactive species. Within the CTRW formalism, the concentration $Q_m(x,t)$ can be expressed as
\begin{eqnarray}
&Q_m(x,t)=\sum_{N=0}^{+\infty}p_N(x)\times \nonumber \\
&\times\left[ \int_{0}^{t} w_N(t')\left( 1-\int_{0}^{t-t'} w(t'') dt''\right)dt' \right],
\end{eqnarray}
where the quantity between square brackets corresponds to ${\cal P}(N|t)$ in Eq. (\ref{PNT}) (see, e.g., \cite{dentz}). Then, integrating $P_m(x,t)$ over time (so to obtain the cumulative distribution $P^c_m(x)$) and computing the $r$-th moment finally leads to expression (\ref{eq_moments}).

Assuming now that the single jump length has a Gaussian distribution with variance $\sigma^2$ and mean $\mu$, then $p_N(x)$ is again a Gaussian distribution with variance $N \sigma^2$ and mean $N \mu$. Therefore, the first and second moment of the cumulative particle distribution respectively read
\begin{eqnarray}
&E[x]= \mu \left\langle N \right\rangle \nonumber \\
&E[x^2]= \sigma^2 \left\langle N \right\rangle + \mu^2 \left\langle N^2 \right\rangle,
\end{eqnarray}
where brackets denote the average with respect to ${\cal P}(N)$. Finally, the radioactive species displacement is provided by the first moment $E[x]$, whereas its spread can be expressed on the basis of the second centered moment $S=E[x^2]-E[x]^2$ \cite{dagan}.
 
Furthermore, the link between $P_m(x,t)$ and $Q_m(x,t)$ allows the moments $E[x^r]$ to be expressed as a function of the memory kernel $M(t)$. Note indeed that $P^c_m(x)$ can be represented in terms of the Laplace transform of $Q_m(x,t)$, namely
\begin{equation}
P^c_m(x)= \bar{Q}_m(x,\lambda).
\end{equation} 
Then, it immediately follows that the moments of $P^c_m(x)$ are given by the Laplace transforms of the moments of $Q_m(x,t)$. General expressions for multidimensional cases are provided, for instance, in \cite{dentz}: in one dimension, we have
\begin{eqnarray}
&E[x]=\mu \lambda^{-1} \bar{M}(\lambda) \nonumber \\
&E[x^2]=\sigma^2 \lambda^{-1} \bar{M}(\lambda)+2\mu^2 \lambda^{-2} \bar{M}(\lambda)^2.
\label{moments_m}
\end{eqnarray}

\section{Normal diffusion}
\label{normal}

We can now specialize this general formalism to the case of normal and anomalous diffusion. Within the CTRW approach, normal diffusion is usually modelled assuming that $w(t)$ is an exponential distribution with mean $\tau$ \cite{rev_geo, cortis_berkowitz}. In this case, the Laplace transform reads $\bar{w}(u)=(1+u\tau)^{-1}$, so that the kernel is simply $\bar{M}(u)=\tau^{-1}$. Moreover, the convolution $w_N(t)$ is known analytically and is given by the Gamma distribution \cite{feller}
\begin{equation}
w_N(t)=\frac{t^{N-1}e^{-\frac{t}{\tau}}}{\tau^N \Gamma(N)},
\end{equation}
whose Laplace transform reads
\begin{equation}
\bar{w}_N(u)=\bar{w}(u)^N=\frac{1}{(1+u \tau)^N}.
\end{equation}
We can therefore obtain ${\cal P}(N)$:
\begin{equation}
{\cal P}(N)= \bar{w}(\lambda)^N-\bar{w}(\lambda)^{N+1}=\frac{\tau/\tau_0}{(1+\tau/\tau_0)^{N+1}},
\label{PN_exp}
\end{equation}
where $\tau_0=\lambda^{-1}$. A numerical example is provided in Fig.~\ref{fig1}, where we compare Eq. (\ref{PN_exp}) with Monte Carlo simulation. For each simulated particle, a random decay time $T$ is first drawn from an exponential pdf with mean $\tau_0$. Then, the particle trajectory is followed until the cumulative waiting time (each contribution being drawn from an exponential pdf with mean $\tau$) exceeds $T$, and the number of performed jumps is recorded. Parameter values are provided in the figure caption. Finally, noting that $\sum_{k=0}^\infty k q^k = q/(1-q)^2$ and $\sum_{k=0}^\infty k^2 q^k = q(1+q)/(1-q)^3$, provided that $|q|<1$, we can compute the moments
\begin{eqnarray}
&E[x]= \frac{\mu}{\tau} \tau_0 \nonumber \\
&E[x^2] \simeq 2 \frac{\sigma^2}{2\tau} \tau_0 + 2 \left(\frac{\mu}{\tau}\right)^2 \tau_0^2.
\label{E2_exp}
\end{eqnarray}
We assume that the time scale of transport is shorter than the time scale of decay ($\tau \ll \tau_0$), hence the approximation sign. In formula (\ref{E2_exp}), $\sigma^2/2\tau$ is the diffusion coefficient $D_m$ and $\mu/\tau$ is the local particle velocity $v_m$ (induced by the forward bias $\mu$) appearing in Eq. (\ref{P_diff}) \cite{cortis_berkowitz}. The same result could be obtained by resorting to expression (\ref{moments_m}) and substituting the specific functional form of $\bar{M}(u)$.

When $w(t)$ is an exponential pdf, the cumulative distribution $W_N(t)$ is known exactly, namely
\begin{equation}
W_N(t)=\frac{\gamma(N,\frac{t}{\tau})}{\Gamma(N)},
\end{equation}
where $\gamma(N,x)=\int_0^{x}s^{N-1}e^{-s}ds$ is the (lower) incomplete Gamma function. Then, we can explicitly compute
\begin{equation}
{\cal P}(N|T)=\frac{\gamma(N,\frac{T}{\tau})}{\Gamma(N)}-\frac{\gamma(N+1,\frac{T}{\tau})}{\Gamma(N+1)}.
\end{equation}
This formula can be simplified by resorting to the properties of the incomplete Gamma function, namely $\gamma(N+1,q)=N\gamma(N,q)-q^Ne^{-q}$ \cite{gradshteyn}. We thus get
\begin{equation}
{\cal P}(N|T)= \frac{ \left( \frac{T}{\tau} \right)^N e^{-\frac{T}{\tau}}}{N!},
\label{PNT_exp}
\end{equation}
which is a Poisson distribution with parameter $T/\tau$, as expected: ${\cal P}(N|T)$ is indeed a counting process for Markovian events whose average rate is $\tau^{-1}$, over a time interval $T$. A numerical example is provided in Fig.~\ref{fig2}, where we compare Eq. (\ref{PNT_exp}) with Monte Carlo simulation, which proceeds as in the previous case, provided that the random decay time is replaced by a fixed threshold $T$. Parameter values are given in the figure caption.

These results can be extended to a broader class of distributions. It can be shown that any waiting time pdf with finite first moment would lead to an expansion of the kind $\bar{w}(u) \simeq 1-c_1 u\tau$ to the first order in $u$, i.e. sufficiently far from the source ($u\tau \ll 1$) \cite{klafter1}. The constant $c_1>0$ depends on the functional form of the pdf. To provide an example, for a Pareto distribution of the kind $w(t)=\alpha \tau^\alpha t^{-1-\alpha}$, with $\alpha>1$, we would have $\bar{w}(u)=1-c_1 u\tau + o(u^\alpha)$, with $c_1=\alpha/(\alpha-1)$. In order to recover normal diffusion, finiteness also of the second moment of the pdf $w(t)$ is required in case of a non vanishing bias $\mu$, which therefore implies $\alpha>2$ \cite{shlesinger, margolin, dentz_berkowitz}. Then, it follows that $\bar{w}(u)^N \simeq 1/(1+c_1 u \tau)^N$ and formulas (\ref{PN_exp}) and (\ref{E2_exp}), which have been derived for the exponential distribution, would remain asymptotically valid, provided that we replace $\tau \to c_1\tau$.

\section{Anomalous diffusion}
\label{subdiffusion}

To illustrate the case of anomalous diffusion, a convenient choice is assuming $\bar{w}(u)=1/(1+(u\tau)^\alpha)$, with $0<\alpha<1$, so that $w(t) \sim t^{-1-\alpha}$, for $t \to \infty$, and the kernel reads $\bar{M}(u)=u^{1-\alpha}/\tau^\alpha$ \cite{rev_geo, klafter1}. The parameter $\tau$ is a characteristic time constant. Then,
\begin{equation}
\bar{w}_N(u)=\bar{w}(u)^N=\frac{1}{(1+(u\tau)^\alpha)^N}
\end{equation}
and we can therefore easily compute ${\cal P}(N)$:
\begin{equation}
{\cal P}(N)=\bar{w}(\lambda)^N-\bar{w}(\lambda)^{N+1}=\frac{(\tau/\tau_0)^\alpha}{(1+(\tau/\tau_0)^\alpha)^{N+1}},
\label{PN_pow}
\end{equation}
where $\tau_0=\lambda^{-1}$ as before. A numerical example is provided in Fig.~\ref{fig3}, where we compare Eq. (\ref{PN_pow}) with Monte Carlo simulation for $\alpha=0.5$. The simulation proceeds similarly as in the case of normal diffusion, the waiting times being now drawn from a power-law pdf. Parameter values are provided in the figure caption. We finally get the moments
\begin{eqnarray}
&E[x]= \frac{\mu}{\tau^\alpha}\tau_0^\alpha \nonumber \\
&E[x^2] \simeq 2 \frac{\sigma^2}{2\tau^\alpha} \tau_0^\alpha +2 \left(\frac{\mu}{\tau^\alpha}\right)^2 \tau_0^{2 \alpha}.
\label{E2_pow}
\end{eqnarray}
Similarly as in the case of normal diffusion, we assume that transport occurs on a time scale shorter than the time scale of decay ($\tau \ll \tau_0$), hence the approximation sign. In formula (\ref{E2_pow}), $\sigma^2/2\tau^\alpha$ can be regarded as the generalized diffusion coefficient $D^\ast_m=\sigma_m^2/2\tau^\alpha$ and $\mu/\tau^\alpha$ as the generalized local particle velocity $v^\ast_m=\mu_m/\tau^\alpha$ implicitly appearing in Eq. (\ref{P_m}) \cite{cortis_berkowitz, klafter1}. This is true for the particular functional form of the memory kernel adopted here. The same result could be obtained by resorting to expression (\ref{moments_m}) and substituting the specific functional form of $\bar{M}(u)$.

In this context, the long time behavior of the reactive species concentration $P_m(x,t)$ can be explicitly obtained. For the case of a vanishing bias ($\mu=0$), note that the contaminant concentration $Q_m(x,t)$ (without radioactive decay) can be expressed in closed form by means of the Fox's $H$ function
\begin{eqnarray}
&Q_m(x,t)=\frac{1}{4 D^\ast_m t^\alpha}\times \nonumber \\
&\times H_{1,1}^{1,0}\left[\frac{|x|}{\sqrt{D^\ast_m t^\alpha} }\Big\vert \begin{array}{c}
(1-\alpha/2,\alpha/2)\\
(0,1)
\end{array}\right], 
\end{eqnarray}
provided that the solution is evaluated sufficiently far from the source \cite{podlubny, klafter1}. The $H$ function admits a computable representation as a series expansion, with an exponentially stretched decay $\log Q_m(x,t) \sim -\left(|x| /t^{\alpha/2}\right) ^{1/(1-\alpha/2)}$ \cite{podlubny, klafter1}. Then, the asymptotic properties of $P_m(x,t)$ immediately follow from $P_m(x,t)= Q_m(x,t)e^{-\lambda t}$.

In some specific cases, analytic results can be obtained for the distribution ${\cal P}(N | T)$. To provide an example, for the L\'evy-Smirnov pdf $w(t)=(\tau/4\pi)^{1/2}e^{-\tau/4t}t^{-3/2}$, which has a power-law decay with $\alpha=0.5$ \cite{feller}, the inverse Laplace transform appearing in Eq. (\ref{PNT}) can be explicitly evaluated, so that ${\cal P}(N | T)$ can be expressed in closed form as
\begin{equation}
{\cal P}(N | T)=\varphi\left(\frac{N+1}{2} \sqrt{\frac{\tau}{T}}\right) -\varphi\left(\frac{N}{2} \sqrt{\frac{\tau}{T}}\right),
\label{PNT_pow}
\end{equation}
where $\varphi(x)=2\pi^{-1/2}\int_0^x e^{-s^2}ds$ is the error function. A numerical example is provided in Fig.~\ref{fig4}, where we compare Eq. (\ref{PNT_pow}) with Monte Carlo simulation. Parameter values are provided in the figure caption. In the general case, ${\cal P}(N | T)$ can be computed from definition (\ref{PNT}) with arbitrary accuracy by resorting to a numerical inverse Laplace transform algorithm \cite{hollenbeck}.

Similarly as for the case of normal diffusion, it can be shown that any pdf with power-law decay and infinite first moment would asymptotically lead to a Laplace transform of the kind $\bar{w}(u) = 1- c_\alpha (u \tau)^\alpha + o(u)$, truncating the expansion to the first non constant term for $u\tau \ll 1$, i.e. evaluating the contaminant concentration sufficiently far from the source \cite{klafter1}. The constant $c_\alpha>0$ depends on the specific details of $w(t)$: for the case of the Pareto pdf, for example, $c_\alpha=\Gamma(1-\alpha)$. The expression of $\bar{w}(u)$ can be regarded as the first order expansion of a pdf $\bar{w}(u) \simeq 1/(1+(u\tau)^\alpha)$. Therefore, formulas (\ref{PN_pow}) and (\ref{E2_pow}) would remain asymptotically valid, provided that we replace $\tau^\alpha \to c_\alpha \tau^\alpha$.

\section{Conclusions}
\label{conclusions}

In this paper we have considered reaction--advection--diffusion processes within the CTRW framework, in both homogeneous and heterogeneous media, the latter giving rise to anomalous transport for the migrating species. We have analytically derived the distribution of the number of jumps that each particle can perform before undergoing a reaction event. On the basis of this result, we have determined the moments of the cumulative particle concentration, which allow the overall displacement and spread of the reacting species to be quantified. Though we have focused on the case of radioactive contaminant particle transport, by virtue of its interest in the field of nuclear waste migration from underground repositories, the proposed framework could be applied to other physical systems where the reaction term is linearly proportional to the concentration of the reacting species, such as first-order chemical reactions.

\acknowledgments
The author would like to express his gratitude to A. Cortis, A. Rosso and two anonymous reviewers for useful discussions and comments. This work has been partially supported by the Italian Ministry of University and Research (MIUR).

\appendix
\section{The memory kernel}
\label{appendix}
Let us briefly recall the definition of the Laplace transform:
\begin{equation}
{\cal L}\left\lbrace g(t)\right\rbrace(u) = \bar{g}(u)= \int_0^{\infty}e^{-ut}g(t)dt.
\end{equation}
The convolution operator ${\cal M}$ is defined as
\begin{equation}
{\cal M}g(t)=\int_0^{+\infty}M(t-t')g(t')dt',
\end{equation}
where the kernel $M(t)$ in the Laplace transformed space satisfies
\begin{equation}
\bar{M}(u)=u\frac{w(u)}{1-w(u)}
\end{equation}
for a sufficiently well behaved function $g(t)$ \cite{rev_geo, cortis_berkowitz}. It immediately follows that
\begin{equation}
{\cal L}\left\lbrace {\cal M}g(t)\right\rbrace =\bar{M}(u)\bar{g}(u).
\end{equation}
The properties of ${\cal M}$ depend on the waiting times distribution $w(t)$. In the direct space, when $w(t)$ has an algebraic decay, $M(t)$ asymptotically behaves as a power-law kernel, accounting for long time correlations: these in turn induce non-Markovian (memory) effects. On the contrary, when $w(t)$ is an exponential distribution the operator reduces to a constant, independent of time, so that the memory effects disappear, the transport process becomes Markovian and normal diffusion is recovered \cite{rev_geo, cortis_berkowitz}.


\begin{thebibliography}{0}
\bibitem{note1} Nevertheless, the CTRW scheme is not universal and many specific realizations of anomalous transport can not be captured within its framework: see, e.g., \cite{havlin} for a comprehensive review and \cite{zoia_kardar} for an example involving diffusion in fracture networks.
\end{thebibliography}
\end{document}